# Acoustic spin pumping as the origin of the long-range spin Seebeck effect

K. Uchida[1,2], H. Adachi[2,3], T. An[1,2], T. Ota[1,2], M. Toda[4], B. Hillebrands[5], S. Maekawa[2,3] & E. Saitoh[1-3,6*]

[1]*Institute for Materials Research, Tohoku University, Sendai 980-8577, Japan*

[2]*CREST, Japan Science and Technology Agency, Sanbancho, Tokyo 102-0075, Japan*

[3]*Advanced Science Research Centre, Japan Atomic Energy Agency, Tokai 319-1195, Japan*

[4]*Graduate School of Engineering, Tohoku University, Sendai 980-8579, Japan*

[5]*Fachbereich Physik and Forschungszentrum OPTIMAS, Technische Universität Kaiserslautern, 67663 Kaiserslautern, Germany*

[6]*PRESTO, Japan Science and Technology Agency, Sanbancho, Tokyo 102-0075, Japan*

[*]e-mail: saitoheiji@imr.tohoku.ac.jp.

**The spin Seebeck effect (SSE) is known as the generation of 'spin voltage' in a magnet as a result of a temperature gradient[1-7]. Spin voltage stands for the potential for spins, which drives a spin current[8-11]. The SSE is of crucial importance in spintronics[12-15] and energy-conversion technology, since it enables simple and versatile generation of spin currents from heat. The SSE has been observed in a variety of materials ranging from magnetic metals[1] and semiconductors[4] to magnetic insulators[3,5]. However, the mechanism, the long-range nature, of the SSE in metals is still to be clarified[2]. Here we found that, using a $Ni_{81}Fe_{19}$/Pt bilayer wire on an insulating sapphire plate, the long-range spin voltage induced by the SSE in magnetic metals is due to phonons. Under a temperature gradient in the sapphire, surprisingly, the voltage generated in the Pt layer is shown to reflect the wire position, although the wire is isolated both electrically and magnetically. This non-local voltage is direct evidence that the SSE is attributed to the coupling of spins and phonons. We demonstrate this coupling by directly injecting sound waves, which realizes the acoustic spin pumping. Our finding opens the door to "acoustic spintronics" in which phonons are exploited for constructing spin-based devices.**

The conventional configuration of the spin Seebeck effect (SSE) in a ferromagnetic metal (FM) is illustrated schematically in Fig. 1a[1-7]. The system comprises a millimetre-sized rectangular FM film and a paramagnetic metal (PM) wire attached to the end of the FM. A uniform temperature gradient, $\nabla T$, is applied to the FM layer along the *x* direction. If a spin voltage is generated in the FM, it injects a spin current with the spatial direction $\mathbf{J}_\mathrm{s}$ and the spin-polarization vector $\boldsymbol{\sigma}$ parallel to the magnetization $\mathbf{M}$ of the FM film into the PM wire (Fig. 1a). This injected spin current is converted into an electric field $\mathbf{E}_\mathrm{ISHE}$ due to the inverse spin Hall effect[16-21] (ISHE) in the PM wire. When $\mathbf{M}$ is along the $\nabla T$ direction, $\mathbf{E}_\mathrm{ISHE}$ is generated along the *y* direction because of the

relation[16,20]

$$\mathbf{E}_{\text{ISHE}} = D_{\text{ISHE}} \mathbf{J}_s \times \boldsymbol{\sigma} \tag{1}$$

The ISHE efficiency $D_{\text{ISHE}}$ is enhanced in noble metals with strong spin-orbit interaction, such as Pt[18,21]. Therefore, by measuring $\mathbf{E}_{\text{ISHE}}$, we can detect the SSE electrically.

The unsolved mystery of the SSE in FM is the long-range spatial distribution of the spin voltage[1], an astonishing feature which has fascinated many researchers. When the FM (e.g. $Ni_{81}Fe_{19}$) /PM system is placed in a temperature gradient (Fig. 1a), the thermally generated spin voltage was found to appear over a millimetre scale in the FM, a situation which appeared to be quite inconsistent with the common knowledge that a spin voltage disappears within a very short distance called the spin-diffusion length[22], typically a few to several hundreds of nanometres in metals.

In this study, we show that the long-range feature of the SSE in FM is, unexpectedly, due to *phonons*. We prepared a sample system consisting of a $Ni_{81}Fe_{19}$/Pt bilayer thin wire placed on a single-crystal insulator sapphire substrate (Fig. 1b). This wire is completely isolated both electrically and magnetically since there are no electric and spin carriers in the sapphire. Owing to this structure, only acoustic vibration, or phonons, can pass through the substrate (Fig. 1d). Therefore, if SSE appears even in this sapphire/[$Ni_{81}Fe_{19}$/Pt-wire] structure, it will become a conclusive proof for the existence of phonon-mediated mechanisms in the SSE in FM. We measured the electric voltage difference $V$ (SSE voltage) between the ends of the Pt layer of the sapphire/[$Ni_{81}Fe_{19}$/Pt-wire] sample at 300 K with controlling the temperature difference $\Delta T$ between the ends of the substrate and with applying an external magnetic field $\mathbf{H}$ (with magnitude $H$) in the $x$-$y$ plane at an angle $\theta$ to the $x$ direction (Fig. 1c).

During the measurements, we confirmed a uniform temperature gradient using an infrared camera. Figure 2a shows a temperature image of the sapphire/[two $Ni_{81}Fe_{19}$/Pt-wires] sample placed in a temperature gradient. As plotted in Fig. 2b, the temperature distribution in the sapphire substrate has a linear profile along the $x$ direction. The inset to Fig. 2b shows that there are no temperature variations along the $y$ direction in the substrate. We also checked that this temperature distribution in the substrate is not changed by attaching voltage probes to the sample (see Supplementary Information (SI) Sec. A).

Figure 2d shows the SSE voltage $V$ as a function of $\Delta T$ at $H$ = 300 Oe in the sapphire/[$Ni_{81}Fe_{19}$/Pt-wire] sample, measured when the $Ni_{81}Fe_{19}$/Pt wire was placed near the lower-temperature end of the substrate. When $\mathbf{H}$ is applied along the $x$ direction ($\theta$ = 0), the $V$ signal was found to appear and the magnitude of $V$ is proportional to $\Delta T$ in the $Ni_{81}Fe_{19}$/Pt wire. In this set-up, Nernst-Ettingshausen effects[23] in the $Ni_{81}Fe_{19}$ and Pt layers are suppressed due to the collinear orientation of $\nabla T$ and $\mathbf{H}$. The inset to Fig. 2e shows that the $V$ signal disappears when the entire sapphire/[$Ni_{81}Fe_{19}$/Pt-wire] sample is uniformly heated to 300 ~ 320 K, confirming again the absence of a heat flow in the $z$ (normal) direction. When **H** is along the $y$ direction ($\theta = 90^\circ$), the $V$ signal also disappears, consistent with equation (1) (Fig. 2d).

Note again that the temperature gradient in the sapphire substrate is uniform and the $Ni_{81}Fe_{19}$/Pt wire is isolated both electrically and magnetically. Nevertheless, as shown in Fig. 2e, the sign of $V$ at finite values of $\Delta T$ is clearly reversed when the $Ni_{81}Fe_{19}$/Pt wire is placed near the higher-temperature end of the substrate. This distinctive behaviour of $V$ can never be explained by the conventional thermoelectric effects since the temperature gradient is of *identical* sign and of same magnitude at every position of the sample (Figs. 2a and 2b).

To further buttress the above results, we measured the magnetic field $H$ dependence of the SSE voltage $V$ for the same sapphire/[$Ni_{81}Fe_{19}$/Pt-wire] system. Figures 2f and 2g show $V$ as functions of $H$ for various values of $\Delta T$ in the $Ni_{81}Fe_{19}$/Pt wire at $\theta = 0$, measured when the wire was placed near the lower- and higher-temperature ends of the substrate, respectively. The sign of $V$ at finite values of $\Delta T$ is reversed by reversing $H$ in both the set-ups, when $|H| > 200$ Oe. This sign reversal of $V$ corresponds to the magnetization **M** reversal of the $Ni_{81}Fe_{19}$ layer, a situation consistent with the aforementioned characteristic properties of the ISHE induced by the SSE[1] through equation (1). We confirmed that the $V$ signal disappears both in the sapphire/[$Ni_{81}Fe_{19}$-wire] sample in which the Pt layer is missing (Fig. 2h) and in the sapphire/[$Ni_{81}Fe_{19}$/Cu-wire] sample in which the Pt layer is replaced by a same-thickness Cu film with weak spin-orbit interaction (Fig. 2i). These measurements confirm that the $V$ signal observed in the sapphire/[$Ni_{81}Fe_{19}$/Pt-wire] sample is attributed to the spin current injected into the Pt layer.

To show how the $V$ signal varies with changing $Ni_{81}Fe_{19}$/Pt-wire position on the sapphire, we attached ten separate $Ni_{81}Fe_{19}$/Pt wires on the sapphire substrate and measured the SSE voltage $V$ in the wires (Fig. 3a). These wires are separated from each other far enough to cut electric and magnetostatic coupling between the wires. As shown in Fig. 3b, in the sapphire/[$Ni_{81}Fe_{19}$/Pt-wire array] sample, $V$ varies linearly with the position of the $Ni_{81}Fe_{19}$/Pt wire; all the results show that the SSE appears even in the $Ni_{81}Fe_{19}$/Pt wire on the insulating sapphire substrate in which the wire is completely isolated electrically and magnetically.

The only possible carrier of the position information in the present system is phonons. Since phonons can pass through even an insulating substrate, the distribution function of magnons in the $Ni_{81}Fe_{19}$ wire is modulated by the nonequilibrium phonons through the magnon-phonon interaction[6,24]. This modulation activates thermal spin pumping[2,3,7] into the Pt layer (Fig. 1d). Since phonons with low frequencies (less than ~ 20 THz: the thermal-phonon-densest frequency at 300 K[25]) exhibit very long propagation lengths[26], magnons in the $Ni_{81}Fe_{19}$ wire can be affected by the substrate temperature at positions far away from the wire, yielding the close to linear dependence of the SSE voltage and the sign reversal of the voltage between the lower- and higher-temperature regions of the sample, which is a characteristic behaviour of SSEs[1-7]. We formulate theoretically the phonon-mediated SSE using a standard many-body technique (see SI Sec. B for details). The SSE voltage in the

present sample structure is described as

$$V = \left(\frac{\theta_{\rm SH}}{\sigma_{\rm c}}\right)\left(\frac{k_{\rm B}e}{w\hbar^3}\right)\tau_{\rm p}\Gamma_{\rm eff}^2 RB\Delta T \tag{2}$$

where $\theta_{\rm SH}$ is the spin-Hall angle of Pt, $\sigma_{\rm c}$ the electric conductivity of Pt, $k_{\rm B}$ the Boltzmann constant, $e$ (>0) the elementary charge, $w$ the width of the Pt wire, $\hbar$ the Planck constant divided by $2\pi$, $\tau_{\rm p}$ the phonon lifetime in the substrate, $\Gamma_{\rm eff}$ the effective magnon-phonon coupling constant near the substrate/Ni$_{81}$Fe$_{19}$ interface, $R$ the strength of the magnetic coupling at the Ni$_{81}$Fe$_{19}$/Pt interface, and $B$ a quantity related to the entropy of the magnon-phonon coupled system (see SI Sec. B), respectively. Equation (2) indicates that the SSE voltage in the present set-up is proportional to $\tau_{\rm p}$ in the substrate and the parameter $\Gamma_{\rm eff}^2$. As also shown in SI Sec. B in detail, $\Gamma_{\rm eff}$ is maximized when the characteristic acoustic impedance[26] $Z \equiv \rho v_{\rm p}$, where $\rho$ is mass density and $v_{\rm p}$ is sound velocity, of the substrate is equal to that of the FM layer.

To confirm this acoustic mechanism further, we measured the voltage $V$ in a glass/[Ni$_{81}$Fe$_{19}$/Pt-wire array] sample where the single-crystal sapphire substrate ($Z \sim 41 \times 10^6$ kg/m$^2$s) is replaced with a same-sized silica-glass substrate ($Z \sim 13 \times 10^6$ kg/m$^2$s) in which the phonon lifetime is very short and the characteristic acoustic impedance is significantly different from that in Ni$_{81}$Fe$_{19}$ ($Z \sim 48 \times 10^6$ kg/m$^2$s). In the glass substrate, the uniform temperature gradient is generated, which is confirmed to be same as that in the sapphire system (compare temperature profiles in Figs. 3d and 3e). Nevertheless, the $V$ signal disappears in the glass/[Ni$_{81}$Fe$_{19}$/Pt-wire array] sample, confirming the important roles of the phonon propagation in the substrate and the acoustic-impedance-matching condition between the substrate and the Ni$_{81}$Fe$_{19}$ layer (compare Figs. 3b and 3c).

Finally, to demonstrate that the aforementioned magnon-phonon coupling allows phonons to generate a spin current, we injected directly a sound wave into a magnet to observe acoustic spin pumping: sound-wave-driven generation of spin currents. To extract the pure contribution of the magnon-phonon interaction, we employed an insulating magnet Y$_3$Fe$_5$O$_{12}$ (YIG) slab as a spin-current injector, where the conduction-electrons' contribution is completely excluded, instead of Ni$_{81}$Fe$_{19}$. Figure 4a shows the set-up for this experiment. The YIG slab is covered with a Pt film and is fixed on a piezoelectric actuator: a polyvinylidene-fluoride (PVDF) film or a lead-zirconate-titanate (PZT) ceramics with different piezoelectric resonance frequencies $f_{\rm p}$ (see the inset to Fig. 4c). By applying an AC voltage with the frequency $f$ (< 10 MHz) across the actuator, we can inject a longitudinal sound wave into the YIG slab. If the sound wave activates magnetic precession via the magnon-phonon coupling, it creates a spin current into the Pt layer and generates a DC voltage along the $x$ direction due to the ISHE in the Pt, when the magnetization **M** of the YIG (∥ **σ** in the Pt) is along the $y$ direction (see equation (1) and Fig. 4b).

Figure 4c shows the $f$ dependence of the DC voltage $V$ between the ends of the Pt layer in the Pt/YIG/PZT samples for various values of the piezoelectric resonance frequency $f_{\rm p}$ of the PZTs, measured with applying the AC voltage to the PZTs and the magnetic field $H$ of 1 kOe along the $y$ direction ($\theta = 90°$). In Fig. 4c, the $f_p$ positions are marked with arrows and the values of $V$ are normalized by the power of the applied AC voltage. In each Pt/YIG/PZT sample, a sharp $V$ dip of negative sign appears clearly around each $f_p$. Since the values of $f_p$ (< 10 MHz) in the present samples are far below ferromagnetic resonance (FMR) frequencies[10] (~ GHz), the $V$ signals observed here are irrelevant to conventional elastically-driven FMR[27-29].

To confirm that the observed $V$ dip is also irrelevant to extrinsic heating of the piezoelectric actuator, we compared the $f$ dependence of $V$ in the Pt/YIG sample with that of the temperature rise of the actuator $\Delta T_p$. To this end, we used the PVDF film instead of the PZT, since the PVDF gives rise to measurable $\Delta T_p$, which monotonically increases with increasing $f$ (Fig. 4e). We found that, in the Pt/YIG/PVDF sample, the $V$ signal of positive sign appears in the higher frequency region ($f > 5$ MHz) and the shape of the $f$-$V$ curve is similar to that of the $f$-$\Delta T_p$ curve (compare Figs. 4d and 4e), indicating that this background of positive sign in the $V$ spectrum is attributed to the conventional SSE due to the heating of the PVDF film, consistent with the previous experiments on the longitudinal SSE in Pt/YIG systems[5]. In contrast, the sharp dip structure of negative sign appears in the Pt/YIG/PVDF sample, although $\Delta T_p$ is negligibly small around the $V$-dip position (Figs. 4d and 4e). Because of this different sign, the $V$-dip signal is irrelevant to the conventional SSE (see SI Sec. C).

Figure 4f shows the $H$ dependence of $V$ in the Pt/YIG sample at the $V$-dip position for various values of the AC-voltage amplitude. When the finite AC voltage is applied, the sign of $V$ is clearly reversed by reversing $H$. We also confirmed that the $V$ signal disappears in the Pt/YIG sample at $\theta = 0$ and in a Cu/YIG sample, where the Pt layer is replaced by a Cu film with weak spin-orbit interaction, consistent with equation (1) (see the inset to Fig. 4d). These results indicate that the observed $V$ dips are due to the ISHE in the Pt layer driven by the acoustic spin pumping, which is direct evidence for the phonon-mediated process in the spin-current generation at far below the FMR frequencies.

The acoustic mechanism observed in the present study can be responsible for the long-range feature of the SSE in metals. For instance, in the previously-reported SSE in conventional $Ni_{81}Fe_{19}$-film/Pt-wire systems[1], the phonon-mediated process through the $Ni_{81}Fe_{19}$ film can explain the long-range spin voltage in the $Ni_{81}Fe_{19}$. The SSE-induced ISHE voltage in the conventional $Ni_{81}Fe_{19}$-film/Pt-wire system is one order of magnitude greater than that observed in the present sapphire/[$Ni_{81}Fe_{19}$/Pt-wire] sample[1], a situation explained by the fact that the path of phonons in the sapphire/[$Ni_{81}Fe_{19}$/Pt-wire] sample is limited to the substrate. The acoustic mechanism provides a new route for collecting heat energy from vast bulk to generate spin and electric voltage. Therefore, the acoustic spin-current generation will develop an innovative design concept for spintronic and energy-saving devices in which phonons are combined with spin currents, paving the way to "acoustic spintronics".

**Acknowledgments**

The authors thank Gerrit E. W. Bauer, Roberto C. Myers, Joseph P. Heremans, Jairo Sinova, Bart J. van Wees and Takahito Ono for valuable discussions. This work was supported by a Grant-in-Aid for Scientific Research in Priority Area 'Creation and control of spin current' (19048009, 19048028), a Grant-in-Aid for Scientific Research A (21244058), the global COE for the 'Materials integration international centre of education and research' all from MEXT, Japan, a Grant for Industrial Technology Research from NEDO, Japan, and Fundamental Research Grants from CREST-JST, PRESTO-JST, TRF, and TUIAREO, Japan.

**Author contributions**

K.U. designed the experiments, fabricated the samples, collected all of the data, and performed analysis of the data. E.S. planned and supervised the study. T.A., T.O. and M.T. support the experiments. H.A. and S.M. developed the explanation of the experiments. K.U., H.A., B.H. and E.S. wrote the manuscript. All authors discussed the results and commented on the manuscript.

**Author information**

The authors declare no competing financial interests. Correspondence and requests for materials should be addressed to E.S.

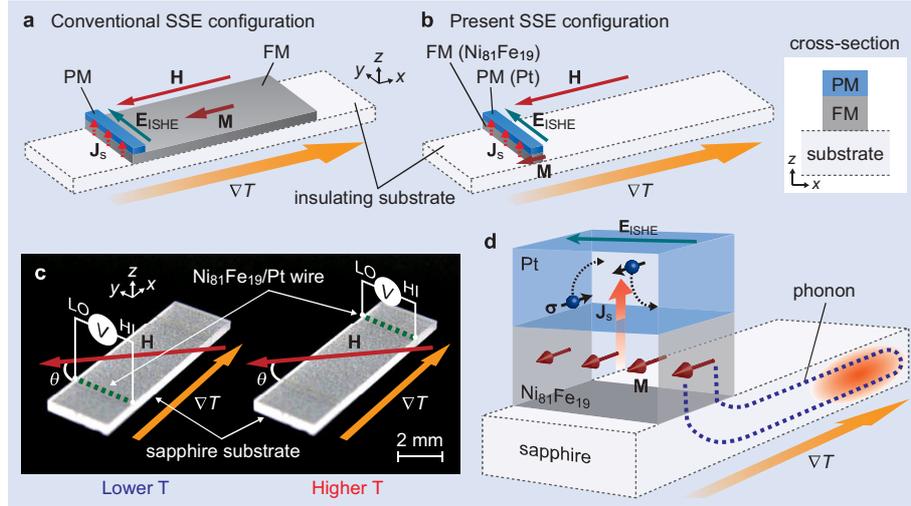

**Figure 1 | Concept of acoustic spin Seebeck effect. a,** Conventional configuration of spin Seebeck effect (SSE). The sample comprises a ferromagnetic metal (FM) film and a paramagnetic metal (PM) wire attached to the end of the FM. When a magnetic field **H** (with magnitude $H$) and a temperature gradient $\nabla T$ are applied along the $x$ direction, an electric field $\mathbf{E}_{\text{ISHE}}$ is generated by the inverse spin Hall effect (ISHE) in the PM wire. **M** and $\mathbf{J}_s$ denote the magnetization of FM and the spatial direction of a spin current, respectively. **b,** Present configuration of SSE. The system comprises an FM ($Ni_{81}Fe_{19}$) /PM (Pt) bilayer wire placed on an insulating substrate. **c,** Photographs of the sample used in the present study. The 20-nm-thick $Ni_{81}Fe_{19}$ layer was deposited on a 0.5-mm-thick sapphire (0001) substrate by electron-beam evaporation, and then the 10-nm-thick Pt layer was sputtered on the $Ni_{81}Fe_{19}$ layer in the same vacuum, yielding clean interfaces. The substrate is of $10 \times 3$ mm$^2$ rectangular shape. The length and width of the $Ni_{81}Fe_{19}$/Pt wire are 3 mm and 0.1 mm, respectively. The $Ni_{81}Fe_{19}$/Pt wire is placed at the distance of 1.8 mm from the end of the substrate. $\theta$ denotes the angle between **H** and the $x$ ($\nabla T$) direction. **d,** The acoustic SSE and the ISHE in the sapphire/[$Ni_{81}Fe_{19}$/Pt-wire] sample. The spin-polarization vector $\boldsymbol{\sigma}$ of the spin current lies along the **M** direction. The dotted line represents phonon propagation.

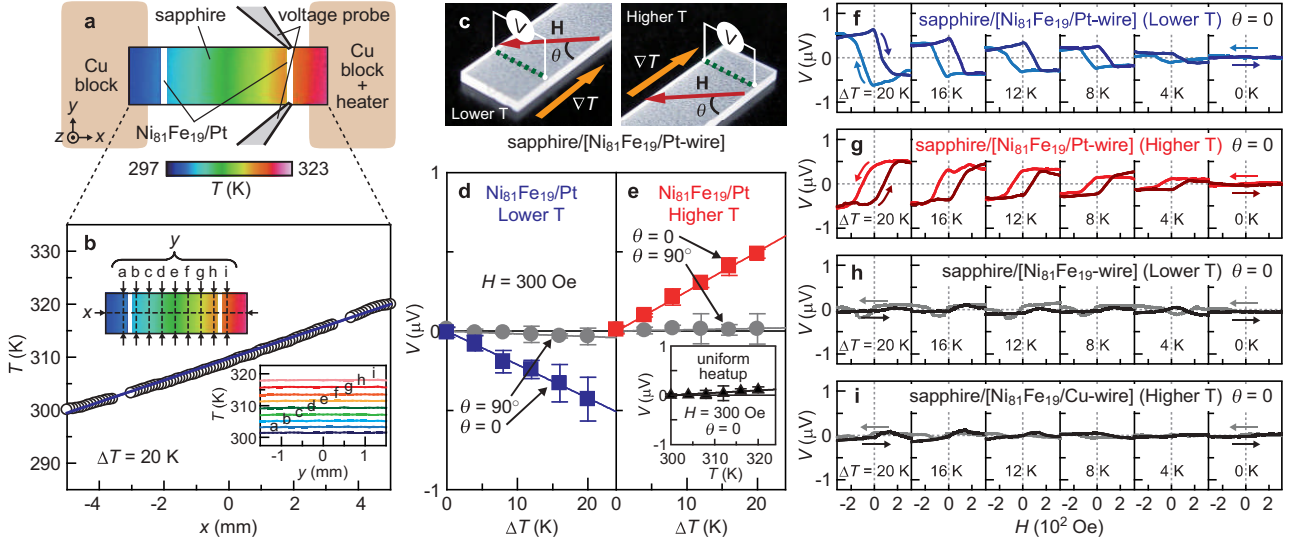

**Figure 2 | Voltage measurement under temperature gradient. a,** Temperature ($T$) image of the sapphire/[two $Ni_{81}Fe_{19}$/Pt-wires] sample, measured with an infrared camera, at $\Delta T = 20$ K. The sample was bridged between two Cu blocks of which temperatures were stabilized to $T = 300$ K and $300$ K $+ \Delta T$. Temperatures on the metallic wires cannot be measured due to very low infrared emittance. **b,** $T$ profile of the sapphire substrate along the $x$ direction at $\Delta T = 20$ K. The inset shows the $T$ profiles along the $y$ direction at various positions (a–i). **c,** Set-ups for the measurement of the acoustic SSE. **d, e,** $\Delta T$ dependence of the voltage $V$ in the sapphire/[$Ni_{81}Fe_{19}$/Pt-wire] sample, measured when the $Ni_{81}Fe_{19}$/Pt wire was placed near the lower-temperature (**d**) and higher-temperature (**e**) ends of the substrate. The magnetic field $H$ of 300 Oe was applied along the $x$ ($\theta = 0$) or $y$ ($\theta = 90°$) directions. The error bars represent 90 % confidence level. The inset shows the $T$ dependence of $V$ in the sapphire/[$Ni_{81}Fe_{19}$/Pt-wire] sample at $\theta = 0$, measured when the entire sample was uniformly heated to $T = 300$–$320$ K. **f, g,** $H$ dependence of $V$ in the sapphire/[$Ni_{81}Fe_{19}$/Pt-wire] sample for various values of $\Delta T$ at $\theta = 0$. **h, i,** $H$ dependence of $V$ in the sapphire/[$Ni_{81}Fe_{19}$-wire] sample (**h**) and in the sapphire/[$Ni_{81}Fe_{19}$/Cu-wire] sample (**i**) for various values of $\Delta T$ at $\theta = 0$, measured when the $Ni_{81}Fe_{19}$ and $Ni_{81}Fe_{19}$/Cu wires were placed near the lower- and higher-temperature ends of the substrate, respectively.

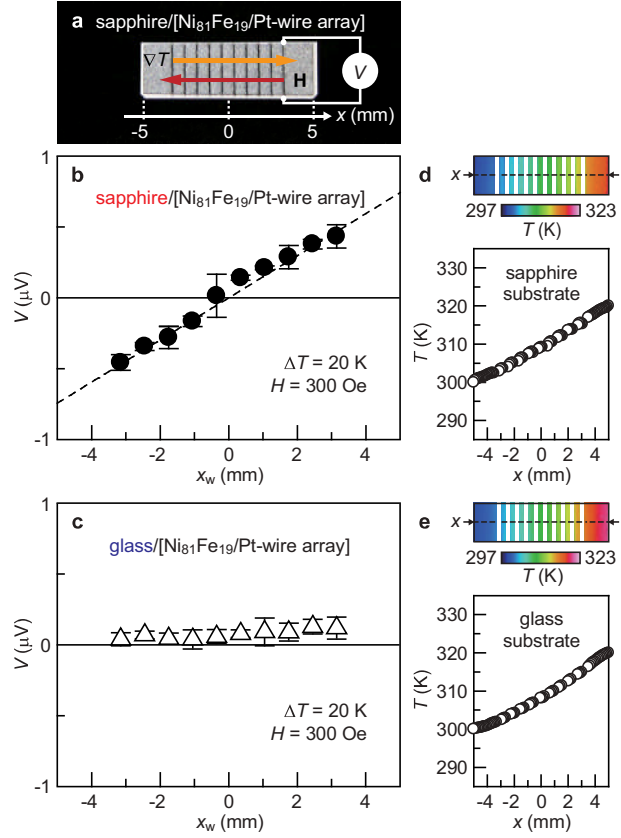

**Figure 3 | Wire-position dependence. a,** Photograph of the sapphire/[Ni$_{81}$Fe$_{19}$/Pt-wire array] sample. The interval between the adjacent Ni$_{81}$Fe$_{19}$/Pt wires is 0.7 mm. **H** and $\nabla T$ were applied along the $x$ direction. **b, c,** $x_w$ dependence of $V$ between the ends of the Ni$_{81}$Fe$_{19}$/Pt wires for the sapphire/[Ni$_{81}$Fe$_{19}$/Pt-wire array] (**b**) and glass/[Ni$_{81}$Fe$_{19}$/Pt-wire array] (**c**) samples at $\Delta T = 20$ K and $H = 300$ Oe. $x_w$ denotes the displacement of the Ni$_{81}$Fe$_{19}$/Pt wire from the centre of the substrate along the $x$ direction. **d, e,** $T$ images and $T$ profiles along the $x$ direction of the sapphire/[Ni$_{81}$Fe$_{19}$/Pt-wire array] (**d**) and glass/[Ni$_{81}$Fe$_{19}$/Pt-wire array] (**e**) samples at $\Delta T = 20$ K.

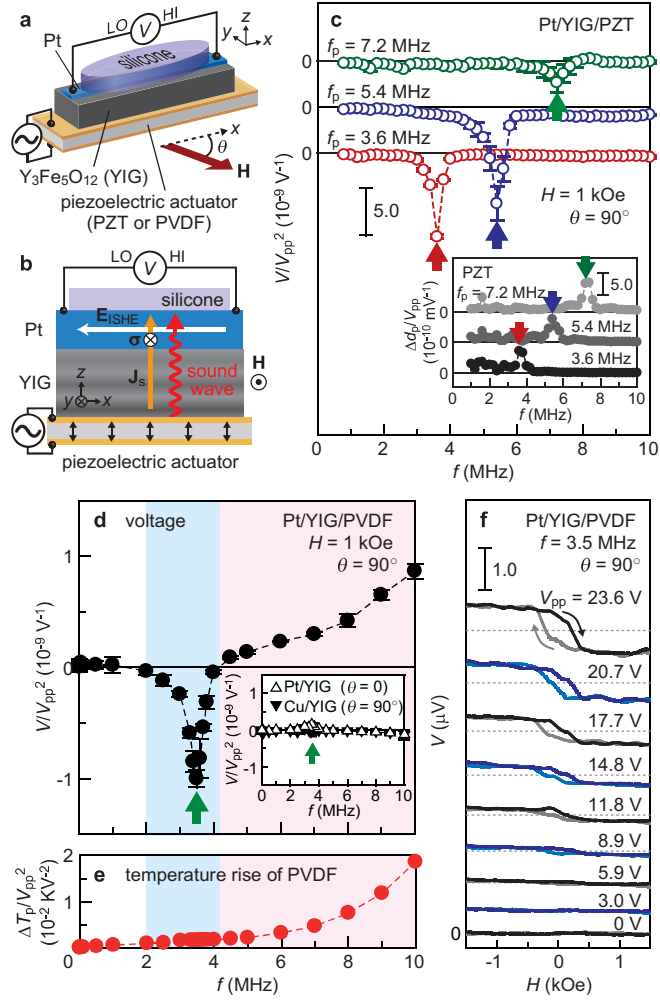

**Figure 4 | Acoustic spin pumping. a,** Experimental set-up. The sample consists of a single-crystal $Y_3Fe_5O_{12}$ (YIG) slab with a Pt film attached to the (100) surface ($x$-$y$ plane) of the YIG. The length, width, and thickness of the YIG slab (Pt film) are 6 mm (6 mm), 2 mm (0.5 mm), and 1 mm (15 nm), respectively. A silicone-rubber heat sink was attached to the top of the Pt/YIG sample. The AC voltage with the frequency $f$ and the peak-to-peak voltage $V_{pp}$ was applied across a piezoelectric actuator: a polyvinylidene-fluoride (PVDF) film or a lead-zirconate-titanate (PZT) ceramics with the piezoelectric resonance frequency $f_p$. **b,** The acoustic spin pumping and the ISHE in the Pt/YIG sample. **c,** $f$ dependence of $V/V_{pp}^2$ in the Pt/YIG/PZT samples for various values of $f_p$ at $H = 1$ kOe and $\theta = 90°$. The inset shows the $f$ dependence of $\Delta d_p/V_{pp}$ in the PZTs with different values of $f_p$, where $\Delta d_p$ denotes the thickness-vibration amplitude of the PZTs, measured by laser Doppler vibrometry. The peak position of each $f$-$(\Delta d_p/V_{pp})$ curve corresponds to $f_p$ of each PZT (marked with arrows). **d,** $f$ dependence of $V/V_{pp}^2$ in the Pt/YIG/PVDF sample at $H = 1$ kOe and $\theta = 90°$. The inset shows the $f$ dependence of $V/V_{pp}^2$ in the Pt/YIG/PVDF ($\theta = 0$) and Cu/YIG/PVDF ($\theta = 90°$) samples. **e,** $f$ dependence of $\Delta T_p/V_{pp}^2$, where $\Delta T_p$ denotes the temperature rise of the PVDF, due to the applied AC voltage, measured with thermocouples. **f,** $H$ dependence of $V$ in the Pt/YIG/PVDF sample for various values of $V_{pp}$ at $f = 3.5$ MHz and $\theta = 90°$. All the measurements were performed at room temperature.

## A. Influence of voltage-probe contact

Here we demonstrate that the contact of voltage probes does not affect the temperature distribution of the sapphire/[$Ni_{81}Fe_{19}$/Pt-wire] samples used in the experiments shown in Figs. 1-3. To check the influence of the voltage-probe contact, we measured temperature $T$ images of the sapphire/[two $Ni_{81}Fe_{19}$/Pt-wires] sample under a temperature gradient $\nabla T$ before and after attaching the voltage probes to the ends of the $Ni_{81}Fe_{19}$/Pt wires (Fig. S1a). As shown also in Fig. S1a, the difference images between them confirm that the temperature of the sapphire/[two $Ni_{81}Fe_{19}$/Pt-wires] sample does not change at every position by attaching the voltage probes (see also the $\Delta T_{\text{diff}}$ profiles along the $x$ ($\nabla T$) direction in Figs. S1b and S1c, where $\Delta T_{\text{diff}}$ denotes the temperature variation of the sample due to the voltage-probe contact).

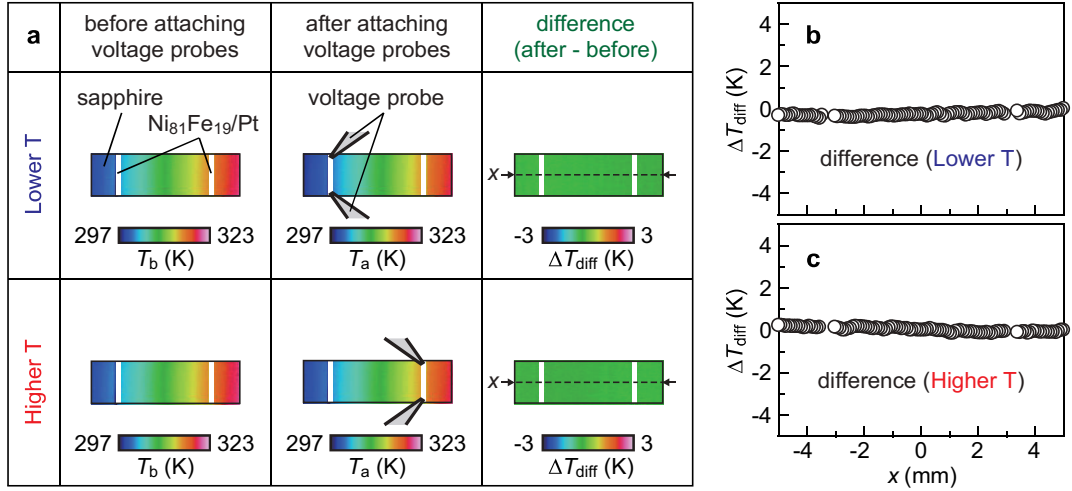

Figure S1 | **Temperature variation due to voltage-probe contact. a,** Temperature images of the sapphire/[two $Ni_{81}Fe_{19}$/Pt-wires] sample, measured with an infrared camera (NEC-Avio TH9100MR). We measured the images before and after attaching the voltage probes (tungsten needles) to the ends of the $Ni_{81}Fe_{19}$/Pt wires with applying the temperature difference of 20 K to the sapphire substrate. Temperatures on the metallic wires cannot be measured due to very low infrared emittance. The images in the right column represent the difference between those in the left and centre columns; $\Delta T_{\text{diff}} = T_a - T_b$, where $T_{b(a)}$ denotes the temperature before (after) attaching the voltage probes to the sample. **b, c,** $\Delta T_{\text{diff}}$ profiles along the $x$ direction, measured when the voltage probes are attached to the ends of the $Ni_{81}Fe_{19}$/Pt wires placed near the lower-temperature (**b**) and higher-temperature (**c**) ends of the substrate.

## B. Linear-response theory of acoustic spin Seebeck effect

In this section, we formulate a linear-response theory of the acoustic spin Seebeck effect, demonstrated in Figs. 1-3 in the main text, and show that the experimental observation can be explained as a consequence of the energy transfer from nonequilibrium phonons to magnons. We consider a model shown in Fig. S2a, and investigate the spin injection from a ferromagnetic metal (FM, in the experiments $Ni_{81}Fe_{19}$) into an attached paramagnetic metal (PM, in the experiments Pt) through the s-d exchange interaction acting at the FM/PM interface:

$$\mathcal{H}_{\rm sd} = J_{\rm sd} \sum_{\bm{r}_0 \in \text{interface}} \bm{s}(\bm{r}_0) \cdot \bm{S}(\bm{r}_0), \qquad (1)$$

where $\bm{s}$ is the conduction-electrons' spin density in PM, $\bm{S}$ is the spin density in FM, and $J_{\rm sd}$ is the strength of the interface s-d exchange coupling. In the spin-wave region, the spin density $\bm{S}$ can be expanded by the magnon creation (annihilation) operator $a_{\bm{q}}^\dagger$ ($a_{\bm{q}}$) with the relation $S^x(\bm{r}_i) = \sqrt{\frac{S_0}{2N_{\rm F}}} \sum_{\bm{q}} (a_{-\bm{q}}^\dagger + a_{\bm{q}}) e^{i\bm{q}\cdot\bm{r}_i}$, $S^y(\bm{r}_i) = -i\sqrt{\frac{S_0}{2N_{\rm F}}} \sum_{\bm{q}} (a_{-\bm{q}}^\dagger - a_{\bm{q}}) e^{i\bm{q}\cdot\bm{r}_i}$, and $S^z(\bm{r}_i) = -S_0 + \frac{1}{N_{\rm F}} \sum_{\bm{q},\bm{Q}} a_{\bm{q}}^\dagger a_{\bm{q}+\bm{Q}} e^{i\bm{Q}\cdot\bm{r}_i}$, where $S_0 = |\bm{S}|$ and $N_{\rm F}$ is the number of lattice sites in FM. The spin current $J_{\rm s}$ injected into PM can be calculated as a rate of change of the spin density in PM as $J_{\rm s} = \frac{\hbar}{2} \sum_{\bm{r} \in N} \langle \partial_t \bm{s}(\bm{r},t) \rangle$, where $\langle \cdots \rangle$ denotes the statistical average at a given time $t$. Assuming that the spin-orbit interaction is weak enough in the neighborhoods of the interface, the Heisenberg equation of motion for $\bm{s}$ yields[7]

$$J_{\rm s} = \sum_{\bm{q},\bm{k}} \frac{-2\mathcal{J}_{\rm sd}^{\bm{k}-\bm{q}} \sqrt{S_0}}{\sqrt{2N_{\rm F} N_{\rm P}}} \int_{-\infty}^{\infty} \Re e C_{\bm{k},\bm{q}}^<(\omega), \qquad (2)$$

where $\mathcal{J}_{\rm sd}^{\bm{k}-\bm{q}}$ is the Fourier transform of $\mathcal{J}_{\rm sd}(\bm{r}) = J_{\rm sd} \sum_{\bm{r}_0 \in \text{interface}} a_{\rm F}^3 \delta(\bm{r} - \bm{r}_0)$ with the effective block spin volume $a_{\rm F}^3$. Here, $C_{\bm{k},\bm{q}}^<(\omega)$ is the Fourier transform of the interface correlation $C_{\bm{k},\bm{q}}^<(t,t') = -i\langle a_{\bm{q}}^+(t') s_{\bm{k}}^-(t) \rangle$ between the magnons and the spin density $s_{\bm{k}}^- = \frac{1}{2\sqrt{N_{\rm P}}} \sum_{\bm{r}_i} [s^x(\bm{r}_i) - is^y(\bm{r}_i)] e^{-i\bm{k}\cdot\bm{r}_i}$ with $N_{\rm P}$ being the number of lattice sites in PM (note that the definition is different from Ref. 7 by a factor $\sqrt{S_0}$).

Using equation (2), it has been shown that, when both the spin density in FM and the conduction electrons in PM are in local thermal equilibrium, no spin current is injected into PM because of the balance between thermal fluctuations in PM and those in FM[7]. Conversely, when the spin density in FM deviates from local thermal equilibrium, a finite spin current is injected into PM. One of such processes was already considered in Ref. 7 where magnons in FM are deviated from local thermal equilibrium by the lateral correlation of magnons feeling the temperature gradient inside FM.

Another thermal spin injection process is nonequilibrium phonons which drive nonequilibrium dynamics of magnons[6]. To discuss such a phonon-mediated spin injection process, we consider the exchange Hamiltonian describing the dynamics of the spin density $\bm{S}$ in FM:

$$\mathcal{H}_{\rm ex} = -\sum_{\bm{R}_i, \bm{R}_j} J_{\rm ex}(\bm{R}_i - \bm{R}_j) \bm{S}(\bm{R}_i) \cdot \bm{S}(\bm{R}_j), \qquad (3)$$

where $J_{\text{ex}}(\boldsymbol{R}_i - \boldsymbol{R}_j)$ is the strength of the exchange coupling between the ions at $\boldsymbol{R}_i$ and $\boldsymbol{R}_j$. The instantaneous position of the ion is written as $\boldsymbol{R}_i = \boldsymbol{r}_i + \boldsymbol{u}(\boldsymbol{r}_i)$ where the lattice displacement $\boldsymbol{u}(\boldsymbol{r}_i)$ is separated from the equilibrium position $\boldsymbol{r}_i$. In the following, we consider a situation where the polarization index $\zeta$ is not mixed with each other, hence $\zeta$ is hereafter omitted. Note that in the case of the acoustic spin pumping, demonstrated in Fig. 4 in the main text, the dominant contribution comes from the longitudinal component because of the selection rule in the magnon-phonon interaction. Up to the linear order in the displacement, the exchange Hamiltonian (3) can be written in the form

$$\mathcal{H}_{\text{ex}} = \sum_{\boldsymbol{q}} \omega_{\boldsymbol{q}} a_{\boldsymbol{q}}^{\dagger} a_{\boldsymbol{q}} + \mathcal{H}_{\text{mag-ph}}, \tag{4}$$

where $\omega_{\boldsymbol{q}} = 2S_0 \sum_{\boldsymbol{\delta}} J_{\text{ex}}(\boldsymbol{\delta}) \sum_{\boldsymbol{q}} \left[1-\cos(\boldsymbol{q}\cdot\boldsymbol{\delta})\right]$ is the magnon frequency with the lattice vector $\boldsymbol{\delta}$ and $\mathcal{H}_{\text{mag-ph}} = \sum_{\boldsymbol{r}_i,\boldsymbol{\delta}} (g\widehat{\boldsymbol{\delta}}/a_{\text{S}}) \cdot \left[\boldsymbol{u}(\boldsymbol{r}_i) - \boldsymbol{u}(\boldsymbol{r}_i+\boldsymbol{\delta})\right] \boldsymbol{S}(\boldsymbol{r}_i)\boldsymbol{S}(\boldsymbol{r}_i+\boldsymbol{\delta})$ is the magnon-phonon interaction[30] with the magnon-phonon coupling $g$ given by $\nabla J_{\text{ex}}(\boldsymbol{\delta}) = (g/|\boldsymbol{\delta}|)\widehat{\boldsymbol{\delta}}$. The displacement $\boldsymbol{u}$ can be expressed in terms of a phonon field as[31] $\boldsymbol{u}(\boldsymbol{r}_i) = \sum_{\boldsymbol{K}} (i\widehat{\boldsymbol{e}}_{\boldsymbol{K}})\sqrt{\frac{\hbar}{2\nu_{\boldsymbol{K}} M_{\text{ion}} N_{\text{F}}}} \left(b_{\boldsymbol{K}} e^{i\boldsymbol{K}\cdot\boldsymbol{r}_i} - b_{\boldsymbol{K}}^{\dagger} e^{-i\boldsymbol{K}\cdot\boldsymbol{r}_i}\right)$, where $M_{\text{ion}}$ is the ion mass and $b_{\boldsymbol{K}}^{\dagger}$ ($b_{\boldsymbol{K}}$) is the phonon creation (annihilation) operator with wavevector $\boldsymbol{K}$, the polarization vector $\widehat{\boldsymbol{e}}_{\boldsymbol{K}}$, and the phonon frequency $\nu_{\boldsymbol{K}}$. These equations yield the following magnon-phonon interaction:

$$\mathcal{H}_{\text{mag-ph}} = \frac{1}{\sqrt{N_{\text{F}}}} \sum_{\boldsymbol{q},\boldsymbol{K}} \Gamma_{\boldsymbol{K},\boldsymbol{q}} B_{\boldsymbol{K}} a_{\boldsymbol{q}+\boldsymbol{K}}^{\dagger} a_{\boldsymbol{q}}, \tag{5}$$

where $B_{\boldsymbol{K}} = b_{\boldsymbol{K}} + b_{-\boldsymbol{K}}^{\dagger}$ is the phonon field operator and the magnon-phonon vertex is given by $\Gamma_{\boldsymbol{K},\boldsymbol{q}} = 2S_0 g \sum_{\boldsymbol{\delta}} \sqrt{\frac{\hbar \nu_{\boldsymbol{K}}}{2 M_{\text{ion}} v_{\text{p}}^2}} (\widehat{\boldsymbol{\delta}} \cdot \widehat{\boldsymbol{e}}_{\boldsymbol{K}})(\widehat{\boldsymbol{\delta}} \cdot \widehat{\boldsymbol{K}}) \left[1 - \cos(\boldsymbol{q} \cdot \boldsymbol{\delta})\right]$ with the sound velocity $v_{\text{p}}$. When the anisotropy in the magnon-phonon coupling is neglected, we obtain an approximate expression

$$\Gamma_{\boldsymbol{K},\boldsymbol{q}} \approx \widetilde{g}(\hbar\omega_{\boldsymbol{q}}) \sqrt{\frac{\hbar \nu_{\boldsymbol{K}}}{2 M_{\text{ion}} v_{\text{p}}^2}}, \tag{6}$$

where $\widetilde{g} = g \sum_{\boldsymbol{\delta}} / 2 J_{\text{ex}}$ is the dimensionless magnon-phonon coupling constant.

With these apparatus, let us consider the phonon-mediated spin injection process shown in Fig. S2a, where nonequilibrium acoustic phonons driven by the temperature gradient in the substrate (sapphire) disturb the distribution function of magnons in FM ($\text{Ni}_{81}\text{Fe}_{19}$), which then results in a finite spin injection into PM (Pt). The spin current injected into PM in this process is given by

$$\begin{aligned} J_{\text{s}} &= -\frac{\sqrt{2}\hbar(J_{\text{sd}}^2 S_0) N_{\text{int}} \widetilde{\nu}_{\text{p}}^2}{2 N_{\text{P}} N_{\text{F}} N_{\text{S}}^2 (\Lambda/b)} \sum_{\boldsymbol{k},\boldsymbol{q},\boldsymbol{K},\boldsymbol{K}',\boldsymbol{K}''} (\Gamma'_{\boldsymbol{K}'',\boldsymbol{q}})^2 |\Omega_{\boldsymbol{K}''-\boldsymbol{K}}|^2 \\ &\quad \times \int_{-\infty}^{\infty} \frac{d\nu}{2\pi} A_{\boldsymbol{k},\boldsymbol{q}}(\nu) |D_{\boldsymbol{K}''}^R(\nu)|^2 |D_{\boldsymbol{K}}^R(\nu)|^2 \text{Im} D_{\boldsymbol{K}'}^R(\nu) \left[ \coth(\tfrac{\hbar\nu}{2k_{\text{B}}T_2}) - \coth(\tfrac{\hbar\nu}{2k_{\text{B}}T_1}) \right], \end{aligned} \tag{7}$$

where $N_{\text{S}}$ is the number of lattice sites in the substrate, $N_{\text{int}}$ is the number of localized spins at the FM/PM interface, $\widetilde{\nu}_{\text{p}}$ is the characteristic frequency corresponding to the phonon high-energy cutoff, $\Omega_{\boldsymbol{K}''-\boldsymbol{K}}$ represents the phonon interaction between the substrate and FM, $D^R_{\boldsymbol{K}}(\nu) = (\nu - \nu_{\boldsymbol{K}} + i/\tau_p)^{-1} - (\nu + \nu_{\boldsymbol{K}} + i/\tau_p)^{-1}$ is the retarded component of the phonon propagator in the substrate with its mode frequency $\nu_{\boldsymbol{K}}$ and phonon lifetime $\tau_p$, $b/\Lambda$ is the lattice constant of the substrate divided by the dimension of the substrate along the temperature gradient. In the above equation, the quantity $A_{\boldsymbol{k},\boldsymbol{q}}(\nu)$ is given by

$$A_{\boldsymbol{k},\boldsymbol{q}}(\nu) = \int_\omega \mathrm{Im}\chi^R_{\boldsymbol{k}}(\omega)\mathrm{Im}X^R_{\boldsymbol{q}-\boldsymbol{K}}(\omega-\nu)|X^R_{\boldsymbol{q}}(\omega)|^2[\coth(\tfrac{\hbar(\omega-\nu)}{2k_B T_1}) - \coth(\tfrac{\hbar\omega}{2k_B T_1})], \qquad (8)$$

where $\chi^R_{\boldsymbol{k}}(\omega) = \chi_P/(1 + \lambda^2_{\mathrm{sf}}k^2 - i\omega\tau_{\mathrm{sf}})$ is the retarded component of the spin-density propagator in PM with $\chi_P$, $\lambda_{\mathrm{sf}}$, and $\tau_{\mathrm{sf}}$ being respectively the paramagnetic susceptibility, the spin diffusion length, and the spin relaxation time and $X^R_{\boldsymbol{q}}(\omega) = (\omega - \widetilde{\omega}_{\boldsymbol{q}} + i\alpha\omega)^{-1}$ is the retarded component of the magnon propagator with $\widetilde{\omega}_{\boldsymbol{q}} = \gamma H_0 + \omega_{\boldsymbol{q}}$ and $\alpha$ being respectively the magnon frequency and the Gilbert damping constant. In equation (7), the prime in $\Gamma'_{\boldsymbol{K}'',\boldsymbol{q}} = \widetilde{g}'(\hbar\omega_{\boldsymbol{q}})\sqrt{\hbar\nu_{\boldsymbol{K}''}/(2M_{\mathrm{ion}}v_p^2)}$ means that the magnon-phonon interaction is limited to the neighborhoods of the substrate/FM interface due to the heat balance condition, since FM in the present situation is not in contact with a heat bath. Making use of $A_{\boldsymbol{k},\boldsymbol{q}}(\nu) \approx -(\chi_P\omega_{\boldsymbol{q}}\tau_{\mathrm{sf}})\coth(\tfrac{\omega_{\boldsymbol{q}}}{2T_1})[\tfrac{1}{\omega_{\boldsymbol{q}}}\mathrm{Im}\tfrac{\chi_{\boldsymbol{k}}(\omega_{\boldsymbol{q}})}{\chi_P}]^2$ and after a bit lengthy calculation, equation (7) is transformed into

$$J_s = \left(\frac{k_B}{2\hbar^2}\right)\tau_p \Gamma^2_{\mathrm{eff}} R \mathcal{B} \Delta T, \qquad (9)$$

where the dimensionless constant $\Gamma_{\mathrm{eff}}$ is given by $\Gamma^2_{\mathrm{eff}} = \left(\frac{\widetilde{g}'^2 \hbar \widetilde{\nu}_p}{M_{\mathrm{ion}}v_p^2}\right)\sum_{\boldsymbol{K}''}\frac{\Omega^2_{\boldsymbol{K}''}}{\nu^2_{\boldsymbol{K}''}+1/\tau_p^2}$, which should be interpreted as a phenomenological parameter representing an effective magnon-phonon coupling near the substrate/FM interface ($\sim \frac{\widetilde{g}'^2 \hbar \widetilde{\nu}_p}{M_{\mathrm{ion}}v_p^2}$) and an effective phonon transmission amplitude between the substrate and FM ($\sim \sum_{\boldsymbol{K}''}\frac{\Omega^2_{\boldsymbol{K}''}}{\nu^2_{\boldsymbol{K}''}+1/\tau_p^2}$). Note that the latter factor is proportional to $T_{\mathrm{S-F}} = 4Z_S Z_F/(Z_S+Z_F)^2$, where $Z_{\mathrm{S(F)}}$ is characteristic acoustic impedance[26] of the substrate (the FM layer); the factor $T_{\mathrm{S-F}}$ ($\leq 1$) is maximized when $Z_S = Z_F$. In equation (9), $R = \frac{0.1 \times J^2_{\mathrm{sd}} S_0 N_{\mathrm{int}} \chi_P}{\pi^2 (\lambda_{\mathrm{sf}}/a)^3 (\Lambda/b)}$ is a factor measuring the strength of the magnetic coupling at the FM/PM interface and $\mathcal{B} = \frac{(T/\widetilde{T}_{\mathrm{m}})^{9/2}}{20\pi^5}\left(\frac{k_B \widetilde{T}_{\mathrm{m}}\tau_{\mathrm{sf}}}{\hbar}\right)^3 \int_0^{\widetilde{T}_{\mathrm{m}}/T} dv \frac{v^{7/2}}{\tanh(v/2)}$ is a function of thermally excited magnons with $\widetilde{T}_{\mathrm{m}}$ being the characteristic temperature corresponding to the magnon high-energy cutoff. Combining equation (9) with equation (1) in the main text, we obtain equation (2) in the main text.

Using $\lambda_{\mathrm{sf}} = 7$ nm, $a = b = 0.3$ nm, $\theta_{\mathrm{SH}} = 0.01$, $\rho = 0.9$ $\mu\Omega$m, $\chi_P = 1 \times 10^{-6}$ cm$^3$/g, $\tau_{\mathrm{sf}} = 1$ ps, $S_0 = 1$, $J_{\mathrm{sd}} = 1$ eV, $\widetilde{T}_{\mathrm{m}} = 850$ K, and $Z_{\mathrm{S(F)}} = 41$ (48) $\times 10^6$ kg/m$^2$s (the resultant $T_{\mathrm{S-F}}$ is $\sim 1$ at the sapphire/Ni$_{81}$Fe$_{19}$ interface), the dimensionless magnon-phonon coupling constant $\widetilde{g}'^2$ times the phonon lifetime $\tau_p$ in the sapphire substrate is roughly estimated as $\widetilde{g}'^2 \tau_p \approx 6 \times 10^{-14}$ s to reproduce the SSE voltage $V/\Delta T \approx 0.02$ $\mu$V/K observed in the experiments (see Figs. 2d and 2e in the main text).

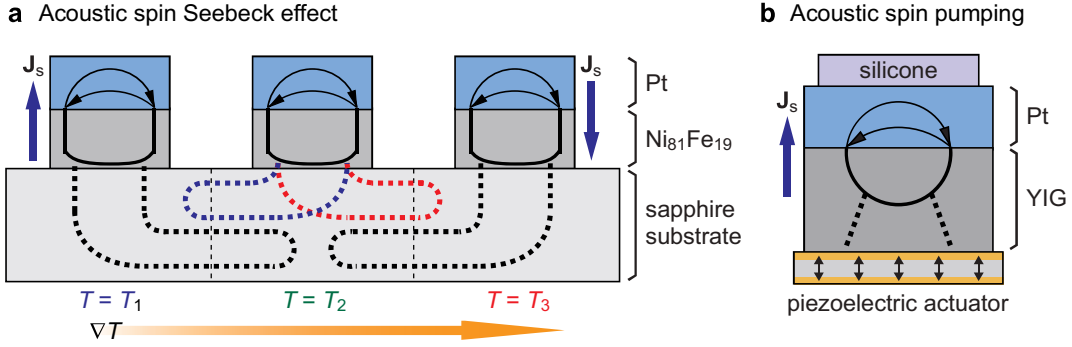

**Figure S2 | Linear-response calculation. a,** Feynman diagrams for calculating spin currents in the Pt layer in the sapphire/[$Ni_{81}Fe_{19}$/Pt-wire] sample used in the experiments on the acoustic spin Seebeck effect (Figs. 1-3 in the main text). Here, the system is divided into three domains with their temperatures $T_1$, $T_2$, and $T_3$. $\mathbf{J}_s$ denotes the spatial direction of the spin current. **b,** Feynman diagram for calculating spin currents in the Pt layer in the Pt/$Y_3Fe_5O_{12}$ (YIG) sample used in the experiments on the acoustic spin pumping (Fig. 4 in the main text). In this case, the interaction between phonons and magnons is given by equation (5) with the thermal phonon operator being replaced by the external sound-wave operator. The thin solid lines with arrows, bold lines, and dotted lines represent electron, magnon, and phonon (or external-sound-wave) propagators, respectively.

## C. Sign difference between acoustic spin pumping and heating effect

Here we discuss the sign of the voltage $V$ signals observed in the Py/$Y_3Fe_5O_{12}$ (YIG) sample, where the signal coming from the sound-wave-driven spin injection is opposite in sign to that coming from the heating effect by the piezoelectric actuator (see Fig. 4c in the main text). In the Pt/YIG system, there are two contributions to the spin injection process; the excitations of magnons in the YIG slab *inject* spin currents into the Pt film (Fig. S2b), while the excitations of conduction electrons in the Pt film *eject* spin currents from itself[7]. Then, in the case of the sound-wave-driven spin injection, the sound wave interacts with magnons in the YIG slab efficiently, while it does not interact with conduction electrons in the Pt film because the thickness of the Pt film (15 nm) is too small for conduction electrons in the Pt film to feel the sound wave ($f = 3.5$ MHz) which has a wavelength of the order of a millimetre. This means that the sound wave *does* excite magnons in YIG slab efficiently while it *does not* excite conduction electrons in the Pt film. That is, spin currents in this case are *injected* into the Pt film (Fig. S2b). On the other hand, in the case of heating by the piezoelectric actuator, the thermal de Broglie length of phonons is much shorter than a millimetre, such that the thermal phonons are able to excite conduction electrons in the Pt film. Moreover, since the silicone rubber in the present set-up acts as a heat sink due to heat conservation, the heat currents carried by thermal phonons are forced to penetrate into the Pt film. Under such circumstances, conduction electrons in the Pt film are excited much stronger than magnons in the YIG slab as the electron-phonon interaction in the Pt film is much stronger than the magnon-phonon interaction in the YIG slab. This means that, in the latter case, spin currents are *ejected* from the Pt film as was demonstrated in recent experiments[5]. These considerations explain the difference in the sign of the signal for both processes as seen in Fig. 4c.

## Additional References